\documentclass[times,twocolumn,final,authoryear]{elsarticle}

\usepackage[english]{babel} 

\usepackage{rotating}

\usepackage{xcolor}
\usepackage{subfig}
\usepackage{graphicx}
\usepackage{csvsimple}
\usepackage{colortbl}
\usepackage{amsmath}
\usepackage{multirow}
\usepackage{hyperref}
\usepackage{svg}
\usepackage{booktabs} 

\usepackage{stfloats} 

\usepackage{ebgaramond}
\usepackage{mwe} 

\usepackage{hyperref} 

\begin{document}

\begin{frontmatter}

\title{Confidence Intervals for Performance Estimates in Brain MRI Segmentation}

\author[1]{Rosana El Jurdi\corref{cor1}}
\author[2]{ Ga\"el Varoquaux}
\author[1]{Olivier Colliot}

\cortext[cor1]{Corresponding author: \href{mailto:eljurdi.rosana@gmail.com}{eljurdi.rosana@gmail.com}}

\address[1]{Sorbonne Universit\'{e}, Institut du Cerveau – Paris Brain Institute - ICM, CNRS, Inria, Inserm, AP-HP, H\^{o}pital de la Piti\'{e}-Salp\^{e}tri\`{e}re, F-75013, Paris, France.}
\address[2]{Soda team, Inria, Saclay, France}

\begin{abstract}
Medical segmentation models are evaluated empirically. As such an
evaluation is based on a limited set of example images, it is
unavoidably noisy. Beyond a mean performance measure, reporting confidence
intervals is thus crucial. However, this is rarely done in medical image
segmentation. The width of the confidence interval depends on the
test set size and on the spread of the performance measure (its
standard-deviation across the test set). For classification, many test
images are needed to avoid wide confidence intervals. Segmentation,
however, has not been studied, and it differs by the amount of
information brought by a given test image.
In this paper, we study the typical confidence intervals in the context of segmentation in 3D brain magnetic resonance imaging (MRI). We carry experiments on using the
standard nnU-net framework, two datasets from the Medical Decathlon
challenge that concern brain MRI (hippocampus and brain tumor segmentation) and two performance measures: the Dice Similarity Coefficient and the
Hausdorff distance. We show that the parametric
confidence intervals are reasonable approximations of the bootstrap
estimates for varying test set sizes and spread of the performance
metric. Importantly, we show that the test size needed to achieve a given
precision is often much lower than for classification tasks. Typically, a 1\%
wide confidence interval requires about 100-200 test samples when the
spread is low (standard-deviation around 3\%). More difficult
segmentation tasks may lead to higher spreads and require over 
1000 samples.
The corresponding code and notebooks are available on GitHub at \href{https://github.com/rosanajurdi/SegVal_Repo}{https://github.com/rosanajurdi/SegVal\_Repo}
\end{abstract}

\end{frontmatter}

\section{Introduction}
 Solid evaluation is crucial for automatic tools used to process
medical images, as medical
professionals must trust the automated system's output.
Most modern segmentation tools rely on deep learning
\cite{hesamian2019deep}, and
their performance depends on details of the training run.
Trained models are evaluated by using unseen data to
estimate their expected performance. Evaluation is harder than it seems.
First, it requires an adequate choice of
metric~\cite{reinke2022metrics,
metrics-reloaded-2,reinke2023understanding}. Second, it is inevitably noisy, 
if only because it uses a finite number of samples.
It is thus crucial to quantify the uncertainty on an estimated performance, for instance using confidence intervals\footnote{Here we use confidence
interval in its broader meaning; in terms of strict statistical
definitions, this encompasses the notion of credible intervals.} (CIs).
Indeed, model developers can easily fall into the trap of apparent model
improvements due to evaluation noise, that do not generalize to new data, as 
in some medical imaging competitions \cite{maier2018rankings,varoquaux2022machine}.

And yet, evaluation uncertainty or CIs are seldom reported in the medical image segmentation
literature. To illustrate how prevalent the problem is, we conducted a survey of all
papers on 3D medical image segmentation published in 2022 in
the following three journals: IEEE Transactions on Medical Imaging,
Medical Image Analysis and Journal of Medical Imaging. As reported in
\tableautorefname~\ref{tab:survey}, 
only 11 of the corresponding 133 paper
reported CIs. Such quick survey, in no way exhaustive,
does suggest that, across three representative journals of the community,
the vast majority ($>90\%$) of recent papers do not report
 confidence intervals for 
 their trained models. Since the initial writing of the present paper, a survey of segmentation papers published at MICCAI (Medical Image Computing and Computer-Assisted Intervention) 2023 has been reported in~\cite{Chr_Confidence_MICCAI2024}. They found that more than 50\% of papers do not assess performance variability at all and that only 0.5\% of papers reported confidence intervals (CIs).
 
Two main factors affect the precision of model evaluation: the size of the test 
set, and how much the performance metric varies across test-set samples. Given
a high spread of the metric across the test set, the precision will be
low and the confidence interval will be wider. Alternatively, increasing
the sample size of the test set leads to tighter confidence intervals. 
In 3D medical image segmentation, the size of the set used to evaluate
the performance is often of small to moderate size, typically in the
order of dozens of subjects, at best hundreds, as obtaining the ground
truth requires voxel-wise annotation by trained raters. One fear is thus
that most segmentation performance assessments will be associated to
large error bars. For image classification, studies have shown
that large sample
sizes are needed for a precise estimation of the prediction accuracy (typically 10,000 samples to achieve a $1\%$-wide confidence interval) \cite{VAROQUAUX201868,varoquaux2022}. Surprisingly, this question has not yet been addressed for medical image segmentation. 

The scope of this paper is that of segmentation of 3D brain magnetic resonance images (MRI). Within this scope, the paper addresses the following questions. What precision can be expected in 3D brain MRI segmentation for typical test set sizes? How trustworthy are the average performance estimates (for instance Dice coefficients or Hausdorff distance) reported in brain MRI segmentation papers? The main objectives are: i) to raise awareness on the importance of reporting confidence interval on independent test sets; ii) to provide the community with typical values of CI that can be expected for various sample sizes and performance variability.

To this end, we present in this paper a series of experiments that deploy nnU-net~\cite{Isensee2021} --a standard framework
for medical image segmentation-- on two classical segmentation tasks from
the Segmentation Decathlon Challenge~\cite{decathlon_short} in order to
estimate confidence intervals that are obtained for test sets of variable
size. The choice of nnUnet was based on its victory in the 2018 challenge, and since then, it has consistently performed comparably to top-ranked models in the live challenge, with little to no difference in performance \cite{decathlon_short}. For each task, we study the following performance measures: the
Dice Similarity Coefficient and the Hausdorff distance. We then report CIs using both
bootstrap (which is the most general approach) and a parametric
estimation, and varying the test set size. We find that the parametric
estimation is in general reasonable, even for small test sets and for
metrics which distribution is far from being Gaussian. Building on this,
we perform simulations for other sizes and spreads. We demonstrate that
the test size needed to achieve a given precision is lower for
segmentation than for classification. This is due to the continuous
nature of the evaluation metrics~\cite{Altman2006}. Indeed, segmentation
algorithms are evaluated with continuous metrics that aggregate multiple
observations at the level of a single image. As each image brings more
information to characterize the errors of the algorithm, tight confidence
intervals can often be obtained with a moderate number of images.

We previously published preliminary work as a conference paper assessing confidence intervals for segmentation~\cite{el2023precise}. The present work differs on many aspects: i) we use a state-of-the-art nnU-net instead of a standard U-Net; ii) we study the Brain Tumor task in addition to the hippocampus; iii) we study both Dice Similarity Coefficient and Hausdorff distance as metrics; iv) we provide an extensive discussion.

The remaining sections of this paper are organized as follows. In section
\ref{sec:CIcomputation}, we describe the statistical tools that we use in
this paper to conduct our analysis. In section \ref{sec: settings}, we
present the datasets examined and the experimental setup. Section
\ref{whole} reports results on the whole test sets. In Section ~\ref{testset-variable}, we then experimentally study how the precision varies when varying the test set size. In Section \ref{sec:simulation}, we provide a table of confidence intervals  when varying the test set sizes and the performance spread (as measured by $\sigma$). Finally, we discuss our findings and conclude in sections~\ref{discussion}~and~\ref{conclusion}.

\begin{table}[t]
    \centering
    \begin{tabular}{|l|cc|c|}
       \hline
       Journal & Nb papers & Nb exp.   & Nb (\%) papers  \\
        &  &   & reporting CI (or SEM) \\
       \hline
         TMI & 51 & 107   & 3 (6\%)  \\
         MedIA & 70 & 171   & 7 (10\%) \\
         JMI & 12 & 22   & 1 (8\%) \\ 
       \hline
        Total & 133 & 300  & 11 (8\%)\\  \hline 
    \end{tabular}
    \caption{\textbf{Papers reporting precision of
segmentation-performance estimates}. Summary of the survey conducted on papers  tackling 3D medical image segmentation via machine/deep learning published in 2022 in the journals IEEE Transactions on Medical Imaging (IEEE TMI), Medical Image Analysis (MedIA) and Journal of Medical Imaging (JMI). We report the number of papers and experiments and the number and proportion of papers that reported the precision of their estimates (we counted both papers reporting confidence intervals (CI) and those reporting standard error of the mean (SEM), since it is very often possible to derive a reasonable estimate of CI from SEM).}
    \label{tab:survey}
\end{table}

\section{Computing Confidence Intervals}
\label{sec:CIcomputation}

In this paper, we are interested in computing confidence intervals for the mean of a given performance metric (e.g. Dice, Hausdorff). We will consider $95\%$ confidence intervals, which are the most common, but everything could be done similarly for other confidence levels.

A general way to compute confidence intervals is through the bootstrap~\cite{efron1994introduction,platt2000bootstrap}. It has the advantage that it does not make any assumption on the distribution of the metric nor on the sampling distribution of the mean. Let us briefly remind the corresponding procedure. Given a test set of size $n$, $M$ bootstrap samples of size $n$ are drawn with replacement from the test set. Each bootstrap sample is denoted as $S^*_m$ where $m \in \{ 1, \ldots, M \}$. In our experiments, we use $M=15000$. We denote $\mu_m^*$\footnote{Throughout the paper, the bootstrap estimate of a given $x$ is always denoted as $x^*$} the mean of $S^*_m$. The $95\% $ confidence interval $[a^*, b^*]$ is the set of values between the $2.5 \% $ and $97.5\%$ percentiles of the sorted bootstrap means $ \{\mu_1^*,\mu_2^*, \ldots, \mu_m^*,\ldots, \mu_M^*\}$. The standard error of the mean $\mu^*$ obtained via bootstrapping ($\text{SEM}^*$) is the standard deviation of the means of all bootstrap samples:
$
\text{SEM}^* = \sqrt{\frac{1}{M} \sum_{m=1}^{M}{\left(\mu_m^*-\mu^*\right)^2}}
$ 
where $\mu^*$ is the mean of the bootstrap sample means $\mu_m^*$.

While the bootstrap is easy to perform, it must be performed by the
researcher evaluating the model, since it requires to have the performance value for each subject of the test set. Knowing typical confidence intervals
to expect given a test size can help evaluating the precision of a result
reported without bootstrap. To build such charts, we perform simulations
to gauge the following classical approximation:
\begin{align}
\begin{split}
 &    \text{SEM} = \frac{\sigma}{\sqrt{n}} \\ 
 &  [a,b] = [\mu-1.96\times \text{SEM},\mu+1.96 \times \text{SEM}]
    \label{eqn:approximation}
\end{split}
\end{align}
where $\mu$ and $\sigma$ respectively denote the mean and standard deviation of the metric over the test set, $n$ the test set size, $\text{SEM}$ the standard error of the mean and $a$ (resp. $b$) are the lower (resp. upper) bounds of the 95\% confidence interval. This approximation relies on the assumption that the sampling distribution of the mean follows a normal distribution. It holds asymptotically with only weak assumptions on the metric distribution (finite expected value and variance) and thus should be valid for sufficiently large samples. However, in practice, it is unknown under which test size this approximation is reasonable. This will be assessed through the experiments in the paper. If the metric distribution is close to normal, the approximation is more likely to be correct even for relatively small samples (the sampling distribution of the mean would thus follow a Student's distribution which becomes quickly close to a normal distribution when $n$ increases). In the following, we will refer to estimations derived from Equations~\ref{eqn:approximation} as {\sl parametric estimations} to distinguish them from {\sl bootstrap estimations}.

Finally, we derive the following quantities from the confidence intervals. First, we are not interested in the absolute values of the boundaries of the confidence interval. Thus, we report confidence intervals independently of the mean, as follows: 
\begin{equation}
    \text{CI} = [a-\mu,b-\mu]
    \label{eqn:CIindependent}
\end{equation}
and we also define the width of $\text{CI}$ as $w=b-a$. When the parametric estimation is used, $w=2\times1.96\times\text{SEM}$. 
 Moreover, different performance measures have different ranges of values
(for instance Dice is between $0\%$ and $100\%$ while the Hausdorff
distance can be arbitrarily large). For the reader to have a better
intuition of the precision of the estimate, we thus normalize the width
by the mean value of the metric. To this end, we define the normalized
width:
\begin{flalign}
    \text{Normalized width}&&\nu = \frac{w}{\mu} &&
    \label{eq:normalized_width}
\end{flalign}
where $\mu$ is the average performance metric mean across the test set. In the case of bootstrap estimation, the above quantities are denoted as $\text{CI}^*$, $w^*$ and $\nu^*$. 

\section{Datasets and Segmentation Method}
\label{sec: settings}

\subsection{Datasets}

We used the Hippocampus and Brain Tumor datasets from the Medical Decathlon challenge~\cite{decathlon_short}. The Hippocampus dataset is mono-modal and composed of 260 3D T1-weighted (T1-w) MPRAGE brain images.
The task is to segment the anterior and posterior parts of the hippocampus (denoted respectively as L1 and L2 according to the challenge annotations).  The Brain Tumor dataset is a multi-modal dataset comprising T1w preconstrast, T1w postcontrast, fluid-attenuated inversion recovery (FLAIR),  and T2-weighted images. It is composed of 484 samples and dedicated to glioma segmentation. The regions to segment are the edema (L1), the non-enhancing tumor (L2), and the enhancing tumor (L3), according to the Decathlon annotations. The data contains a subset of the data used in the 2016 and 2017 Brain Tumor Segmentation (BraTS) challenges~\cite{brats}. The brain tumor task is considered difficult (best Dice is around $70\%$) while the Hippocampus is easier in comparison (best Dice around $90\%$).

From the 260 and 484 samples of the Hippocampus and Brain Tumor dataset respectively, 100 patients were randomly selected for training, 50 for validation and the remaining samples constituted the test set. The size of the training and validation subsets were the same for both datasets to ensure that the number of training samples has no impact on the model performances. 

\subsection{Segmentation Methods}

We used the Decathlon Challenge winner's nnU-net, as a framework to perform our experiments. nnU-net~\cite{Isensee2021} is a fully automated segmentation framework that makes use of typical 2D and 3D U-Net architectures by dynamically and automatically adapting the hyper-parameters to a particular dataset. Given an arbitrary dataset, the nnU-net pipeline first extracts data fingerprints (e.g. voxel spacing, intensity distribution, modalities, class ratios etc.). 
Pre-processing of the datasets is conducted automatically via cropping to non-zero volumes to reduce the computational burden, resampling relative to voxel spacing to allow spatial semantics, and z-score normalization, if need be. 
We allowed nnU-Net to automatically determine the best configuration of parameters. In nnU-net, one can either predetermine which architecture to train on the given dataset or allow nnU-net to choose the best configuration. In the present paper, we conducted our experiments using the  3D full-resolution U-Net and the 2D U-Net. It is possible that even better performances could have been obtained using the cascade approach or letting nnU-net choose the best approach. However, the aim of the present paper is not achieve the highest possible performance but to obtain a set of results which are representative of the state-of-the-art.
All networks are trained via a combined Dice and cross-entropy loss and the Adam optimizer. 
Further information concerning model parameters and the training scheme can be found in~\cite{Isensee2021}.

\subsection{Performance Metrics and Setup}
We have conducted our study for two common performance metrics: the Dice Similarity Coefficient computed for each class separately and the $95\%$ Hausdorff distance\footnote[1]{Performance metrics were computed using this code: \href{https://github.com/deepmind/surface-distance}{\text{https://github.com/deepmind/surface-distance}}}.
For sake of conciseness, we only reported results for the  first region of each task (L1). In the Hippocampus dataset, it is the anterior part. In the Brain Tumor dataset, it corresponds to the edema. Note nevertheless that the training was done simultaneously on all regions. The full code in Python and Jupyter notebooks to reproduce our experiments are available on GitHub\footnote[2]{\href{https://github.com/rosanajurdi/SegVal_Repo}{https://github.com/rosanajurdi/SegVal\_Repo}}



\begin{figure}[b!]
\subfloat[][Hippocampus dataset, 3D U-net]
  {
 	\begin{minipage}[c]{
 	   0.23\textwidth}
 	   \centering
	   \includegraphics[width=4.9cm,height=4.9cm]{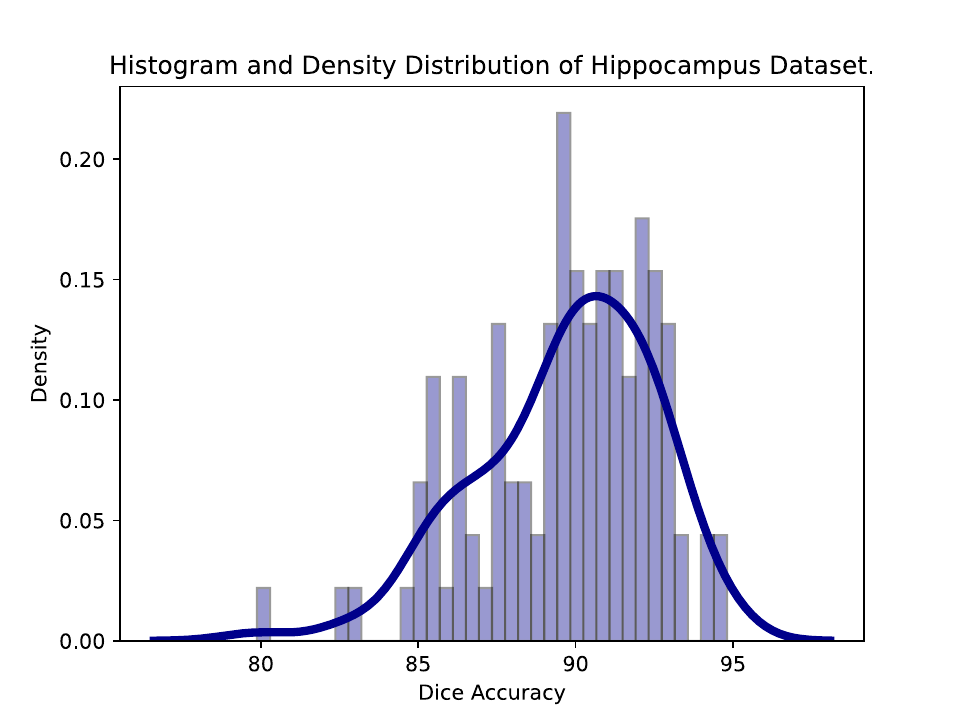} 
	   \label{FA-3D-HIP-DSC}
 	\end{minipage}
 	} 
  \hfill 
\subfloat[][Hippocampus dataset, 2D U-net]
  {
 	\begin{minipage}[c]{
 	   0.23\textwidth}
 	   \centering
	   \includegraphics[width=4.9cm,height=4.9cm]{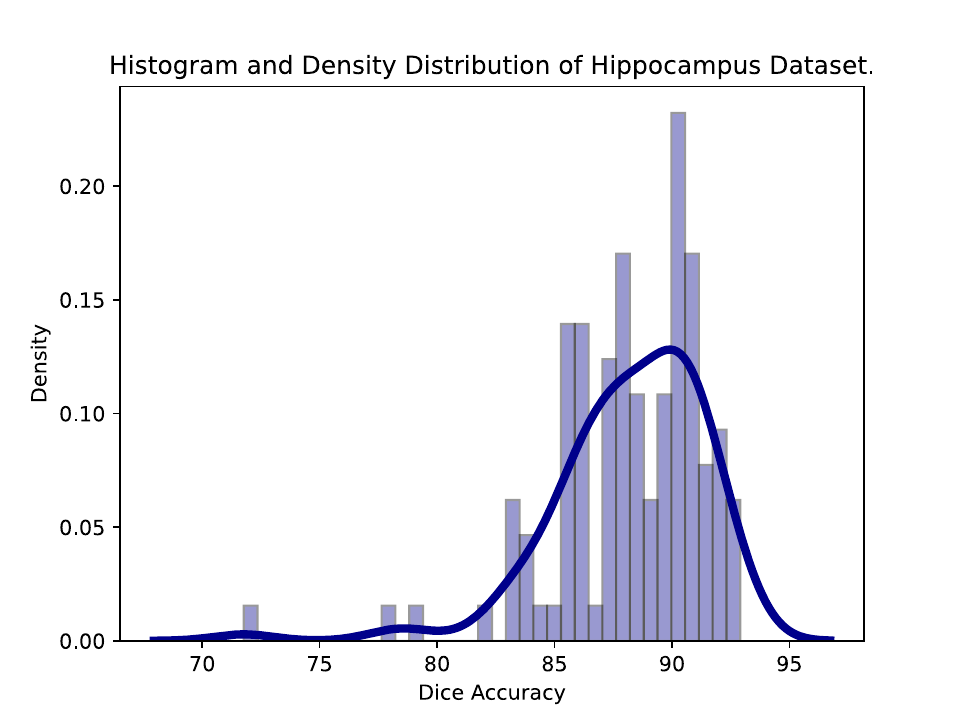} 
	   \label{FA-2D-HIP-DSC}
 	\end{minipage}
 	} 

   \subfloat[][Brain Tumor dataset, 3D U-net ]
  {
 	\begin{minipage}[c]{
 	   0.23\textwidth}
 	   \centering
	   \includegraphics[width=4.9cm,height=4.9cm]{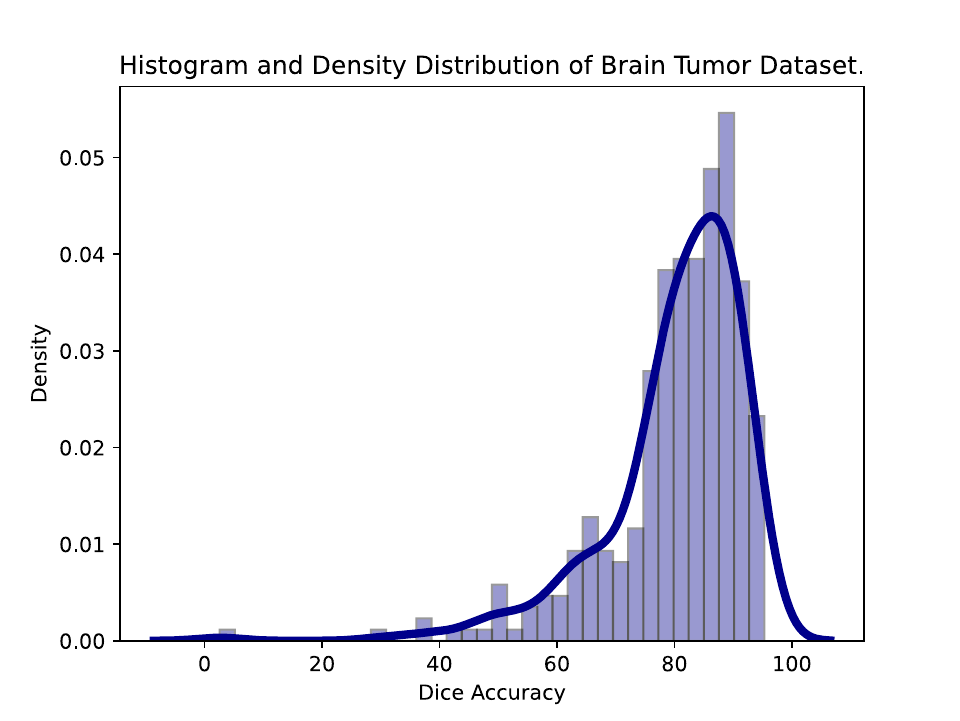} 
	   \label{brain-3d-dsc}
 	\end{minipage}
 	}
   \hfill 
   \subfloat[][Brain Tumor dataset, 2D U-net]
   {
 	\begin{minipage}[c]{
 	   0.23\textwidth}
 	   \centering	   \includegraphics[width=4.9cm,height=4.9cm]{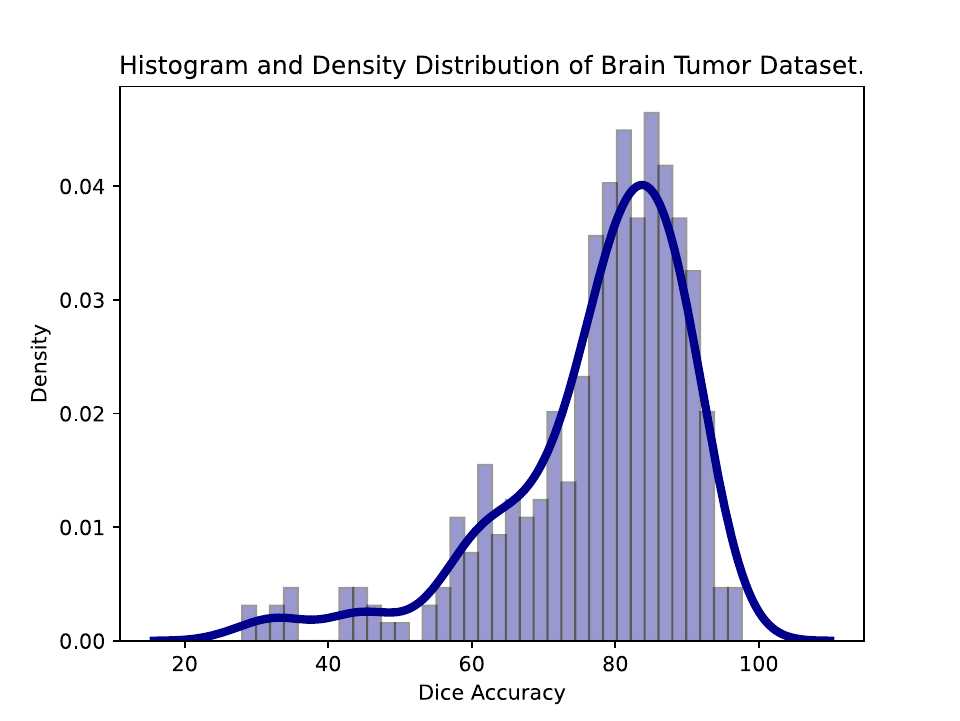} \label{Brain-2d-dsc} 
 	    \\
 	\end{minipage}
 	}
  \hfill 
\caption{\textbf{Histogram of Dice Similarity Coefficient over the entire test set}, shown together with a kernel density estimation (KDE) which smoothes the observations with a Gaussian kernel. (a) Hippocampus dataset, 3D U-Net. (b) Hippocampus dataset, 2D U-Net. (c) Brain Tumor dataset, 3D U-Net. (d) Brain Tumor dataset, 2D U-Net. }
\label{Dicedist1}
\end{figure}



\begin{figure}[b!]%
\hspace*{-1ex}%
\subfloat[][Hippocampus dataset, 3D U-net]
  {
 	\begin{minipage}[c]{
 	   0.23\textwidth}
 	   \centering
	   \includegraphics[width=4.7cm,height=4.9cm]{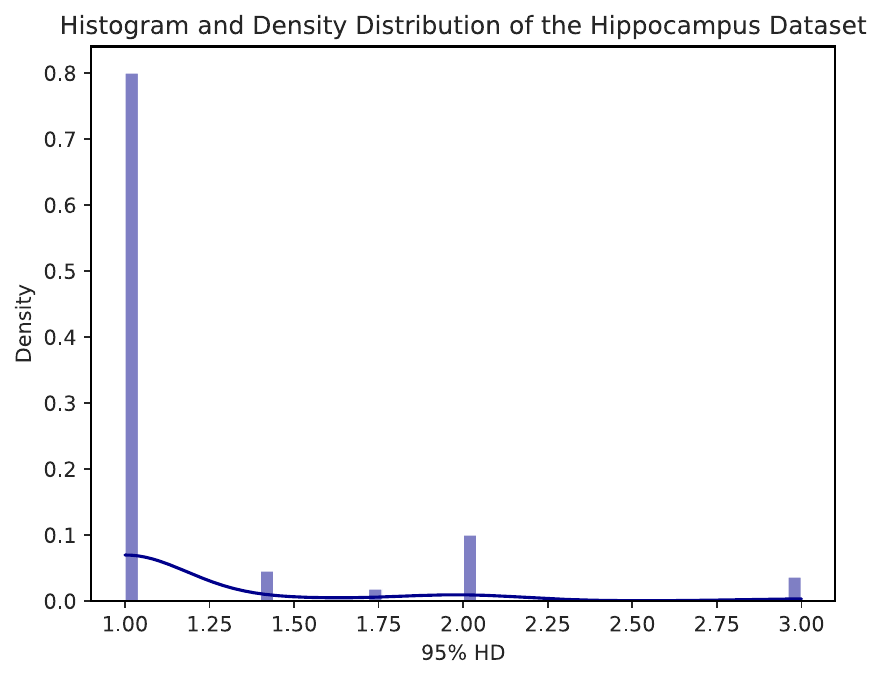} 
	   \label{FA-HIP-HD3D}
 	\end{minipage}
 	} 	
    \hfill 
\subfloat[][Hippocampus dataset, 2D U-net]
  {
 	\begin{minipage}[c]{
 	   0.23\textwidth}
 	   \centering
	   \includegraphics[width=4.7cm,height=4.9cm]{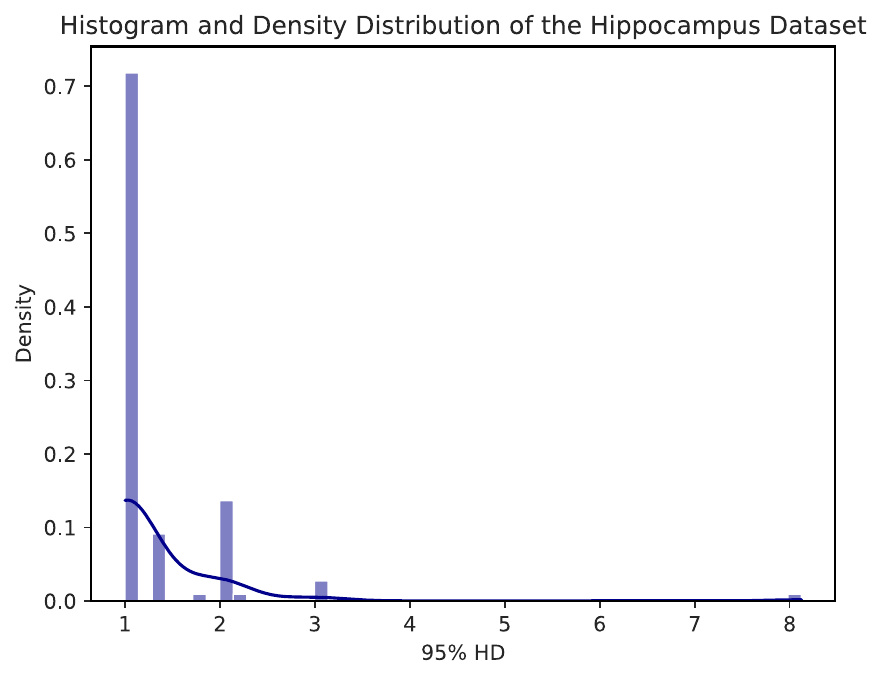} 
	   \label{FA-HIP-HD2D}
 	\end{minipage}
 	} 

\hspace*{-1ex}%
  \subfloat[][Brain Tumor dataset, 3D U-net ]
  {
 	\begin{minipage}[c]{
 	   0.23\textwidth}
 	   \centering
	   \includegraphics[width=4.9cm,height=4.9cm]{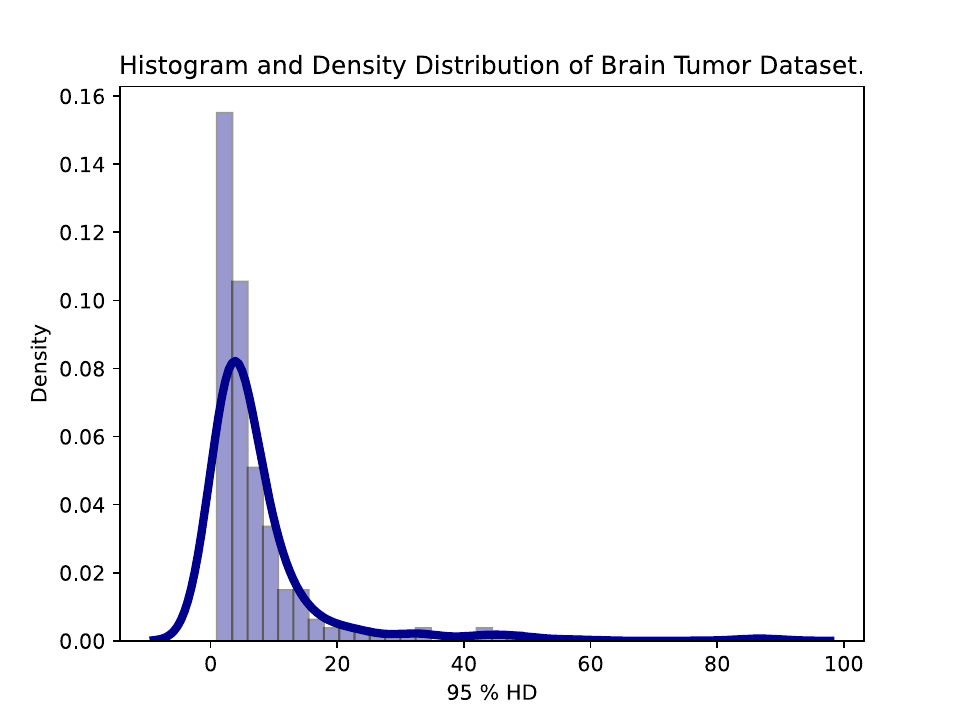} 
	   \label{brain-3d-HD}
 	\end{minipage}
 	}
     \hfill 
   \subfloat[][Brain Tumor dataset, 2D U-net]
   {
 	\begin{minipage}[c]{
 	   0.23\textwidth}
 	   \centering
 	   \includegraphics[width=4.9cm,height=4.9cm]{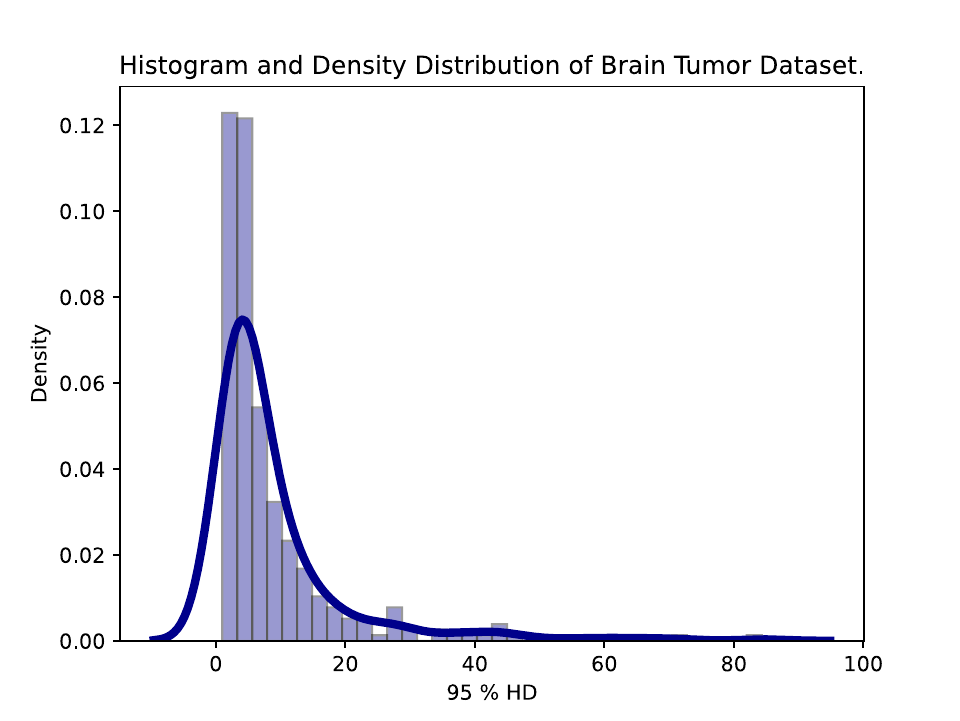} \label{Brain-2d-HD} 
 	    \\
 	\end{minipage}
 	}
\caption{\textbf{Histogram of 95\% Hausdorff distance over the entire
test set}, shown together with a kernel density estimation (KDE) which smoothes the observations with a Gaussian kernel. (a) Hippocampus dataset, 3D U-Net. (b) Hippocampus dataset, 2D U-Net. (c) Brain Tumor dataset, 3D U-Net. (d) Brain Tumor dataset, 2D U-Net.}
\label{haussdist2}
\end{figure}

\begin{table*}[t]
 \centering

\resizebox{\textwidth}{!}{
    \begin{tabular}{|c|cc|ccccc|cccc|}
    \hline

    \hline
& & & \multicolumn{5}{c|}{Parametric} & \multicolumn{4}{c|}{Bootstrap} \\
& & & $\mu$ & $\sigma$ & $SEM$ & $CI$ & $\nu$ & $\mu^*$ & $SEM^*$ & $CI^*$ & $\nu^*$\\ \hline
\multirow{2}{*}{Hippocampus}

&\multirow{2}{*}{3D U-Net} & Dice & 89.714  &  2.797  &  0.267  &  [-0.52, 0.52]  & 0.012 & 
89.715 &  0.263 &  [-0.53, 0.51]
 & 0.012\\
&& 95 \% HD &1.205  &  0.472  &  0.045  &  [-0.08, 0.08] & 0.149 & 1.205 &  0.045 &  [-0.08, 0.09] & 0.149
\\

\cline{2-12}

\multirow{2}{*}{($n=110$)}
& \multirow{2}{*}{2D U-Net} &Dice & 88.197  &  3.267  &  0.311  &  [-0.61, 0.61] & 0.014 & 
88.199 &  0.313 &  [-0.64, 0.59]
 & 0.014\\
&&95\% HD & 1.311  &  0.806  &  0.077  &  [-0.15, 0.15]  & 0.229 &
 1.31 &  0.077 &  [-0.13, 0.17] & 0.229
\\

\hline

\multirow{2}{*}{Brain Tumor }

& {\multirow{2}{*}{3D U-Net}}& Dice &  80.265  &  11.947  &  0.654  &  [-1.28, 1.28]  & 0.032
& 80.268 &  0.659 &  [-1.31, 1.24]
& 0.032\\
& & 95 \% HD & 7.726  &  10.634  &  0.582  &  [-1.14, 1.14]  & 0.294 & 7.73 &  0.581 &  [-1.08, 1.18] & 0.294
\\

\cline{2-12}
\multirow{2}{*}{($n=334$)}
& \multirow{2}{*}{2D U-Net} & Dice  & $77.489$  & $13.115 $ & $0.718$  & [-1.41, 1.41] & 0.036& $77.488$  & $0.717$ & [-1.43, 1.38] & 0.036 \\ 
& & 95 \% HD & 8.855  &  11.262  &  0.616  &  [-1.21, 1.21] & 0.272
& 8.862 &  0.62 &  [-1.22, 1.22] & 0.272\\

\hline

    \end{tabular}
}
    \vspace{0.2cm}
    \caption{\textbf{Results on the full test set ($ n = 110$ for
Hippocampus and $n=334$ for Brain Tumor) for region $L_1$.}
$\mu$~and~$\sigma$  are the empirical mean and standard deviation of the
Dice Similarity Coefficient (Dice) and 95 \% volumetric Hausdorff distance (HD) across
all patients in the test set. $SEM$ is the standard error of the mean, $CI$ is the $95\%$ confidence interval independent of the mean
calculated using the parametric estimation, $\nu$ is the normalized CI. $\mu^*$, $SEM^*$, $CI^*$ and $\nu^*$
are the bootstrap estimates.}
    \label{tab:Stats_on_full_ds}
\end{table*}

\section{Experimental Study on the Whole Test Sets}\label{whole}

In this section, we study the  precision of the performance evaluation using the full test sets (test set sizes are 110 for the Hippocampus and 334 for the Brain Tumor, respectively). 

The distribution of Dice Similarity Coefficient and Hausdorff distance values over the
test set are shown in~\figureautorefname~\ref{Dicedist1}. One can observe
that the Dice Similarity Coefficient's distribution is not very far from Gaussian
despite outliers and skewness in the histogram plots. It is not the case
for the Hausdorff distance (Figure~\ref{haussdist2}). First, this measure
is lower-bounded by one and a lot of 
values lie to one, which makes the distribution highly skewed. Also,
due to the discrete nature of the images, some values will never occur
hence histograms are sparse. 

We then compute, for both metrics and for both 2D and 3D networks, the test sample mean $\mu$, the test sample standard deviation $\sigma$,  $\text{SEM}$, $\text{CI}$ and $\nu$. The computations are done as described in Section~\ref{sec:CIcomputation}, using both parametric and bootstrap estimations. Bootstrapping on the entire dataset took around $0.53$ second for the hippocampus and $0.62$ second for the Brain Tumor dataset on a standard laptop.
Results are reported in~\tableautorefname~\ref{tab:Stats_on_full_ds}. One can see that, under such test set sizes, the computations using parametric estimation are close to those obtained with the bootstrap, even for the Hausdorff distance.  This indicates that it is a reasonable approximation under this size regime. In the next section, we will study if this still holds for smaller test set sizes.

\section{Experimental Study on Sub-samples of the Test Sets }
\label{testset-variable}

In this section, we study experimentally the relationship between the test set size and the precision of the estimation of the segmentation performance. To that end, we draw subsamples of variable size
$$ k \in K=\{10,20,30,50,100,\ldots,n\}$$ where $n$ is the whole test set size. For the Hippocampus dataset $n=110$  and $K=\{10,20,30,50,100,110\}$ and for the Brain Tumor dataset, $n=334$ and $K=\{10,20,30,50,100,200,300,334\}$. 

For small subsamples, the estimates can vary substantially from one
drawing to another. In order not to depend on a particular drawing, which
may be lucky or unlucky, we repeat the procedure 100 times for each value
of $k$. We denote the subsamples as $(S_{k,j})$ where $k$ is the
subsample size and $ j \in \{1, \ldots , 100\}$ is the index of a
particular drawing. We then compute the  precision measures
($\text{SEM}$, $\text{CI}$, $\nu$ and bootstrap counterparts) for the subsamples of different
sizes. The exact formulas for the estimation of the different measures
are detailed in appendix \ref{sec:details_subsample}.

\begin{table*}[b!]
 \centering

 \begin{turn}{90}
\resizebox{\textheight}{!}{%
\begin{tabular}{|lc|cc|cc|cc|cc|cc|ccc|}
\hline
\cellcolor{gray}
$\sigma \downarrow $  & $k \rightarrow$  & 10 & 20 & 30 & 50 & 100 & 200 & 300 & 500 & 1000 & 1500 & 2000 & 2500 & 3000\\  \hline
\multirow{2}{*}{0.47} & $SEM$  & 0.15 & 0.11 & 0.09 &0.07 &0.05 &0.03 &0.03 &0.02 &0.01 &0.01 &0.01 &0.01 &0.01\\  
 & $CI$ &  [-0.29, 0.29] &[-0.21, 0.21] &[-0.17, 0.17] &[-0.13, 0.13] &[-0.09, 0.09] &[-0.07, 0.07] &[-0.05, 0.05] &[-0.04, 0.04] &[-0.03, 0.03] &[-0.02, 0.02] &[-0.02, 0.02] &[-0.02, 0.02] &[-0.02, 0.02]\\ \hline 
 \multirow{2}{*}{0.81} & $SEM$  & 0.26 &0.18 &0.15 &0.11 &0.08 &0.06 &0.05 &0.04 &0.03 &0.02 &0.02 &0.02 &0.01\\  
 & $CI$ &  [-0.5, 0.5] &[-0.35, 0.35] &[-0.29, 0.29] &[-0.22, 0.22] &[-0.16, 0.16] &[-0.11, 0.11] &[-0.09, 0.09] &[-0.07, 0.07] &[-0.05, 0.05] &[-0.04, 0.04] &[-0.04, 0.04] &[-0.03, 0.03] &[-0.03, 0.03]\\ \hline 
 \multirow{2}{*}{1} & $SEM$  &0.32 & 0.22 &0.18 &0.14 &0.1 &0.07 &0.06 &0.04 &0.03 &0.03 &0.02 &0.02 &0.02\\  
 & $CI$ &[-0.62, 0.62] & [-0.44, 0.44] &[-0.36, 0.36] &[-0.28, 0.28] &[-0.2, 0.2] &[-0.14, 0.14] &[-0.11, 0.11] &[-0.09, 0.09] &[-0.06, 0.06] &[-0.05, 0.05] &[-0.04, 0.04] &[-0.04, 0.04] &[-0.04, 0.04]\\ \hline 
 \multirow{2}{*}{2.79} & $SEM$  &0.88 &0.62 &0.51 &  0.39 & 0.28 &0.2 &0.16 &0.12 &0.09 &0.07 &0.06 &0.06 &0.05\\  
 & $CI$ &[-1.73, 1.73] &[-1.22, 1.22] &[-1.0, 1.0] &[-0.77, 0.77] &[-0.55, 0.55] & [-0.39, 0.39] &[-0.32, 0.32] &[-0.24, 0.24] &[-0.17, 0.17] &[-0.14, 0.14] &[-0.12, 0.12] &[-0.11, 0.11] &[-0.1, 0.1]\\ \hline 
  \multirow{2}{*}{3.26} & $SEM$  &1.03 &0.73 &0.6 &0.46 &0.33 & 0.23 &0.19 &0.15 &0.1 &0.08 &0.07 &0.07 &0.06\\  
 & $CI$ &[-2.02, 2.02] &[-1.43, 1.43] &[-1.17, 1.17] &[-0.9, 0.9] &[-0.64, 0.64] & [-0.45, 0.45] &[-0.37, 0.37] &[-0.29, 0.29] &[-0.2, 0.2] &[-0.16, 0.16] &[-0.14, 0.14] &[-0.13, 0.13] &[-0.12, 0.12]\\ \hline 
 \multirow{2}{*}{5} & $SEM$  &1.58 &1.12 &0.91 &0.71 &0.5 &0.35 & 0.29 & 0.22 &0.16 &0.13 &0.11 &0.1 &0.09\\  
 & $CI$ &[-3.1, 3.1] &[-2.19, 2.19] &[-1.79, 1.79] &[-1.39, 1.39] &[-0.98, 0.98] &[-0.69, 0.69] &[-0.57, 0.57] & [-0.44, 0.44] &[-0.31, 0.31] &[-0.25, 0.25] &[-0.22, 0.22] &[-0.2, 0.2] &[-0.18, 0.18]\\ \hline 
  \multirow{2}{*}{10.63} & $SEM$  &3.36 &2.38 &1.94 &1.5 &1.06 &0.75 &0.61 &0.48 &0.34 & 0.27 & 0.24 &0.21 &0.19\\  
 & $CI$ &[-6.59, 6.59] &[-4.66, 4.66] &[-3.8, 3.8] &[-2.95, 2.95] &[-2.08, 2.08] &[-1.47, 1.47] &[-1.2, 1.2] &[-0.93, 0.93] &[-0.66, 0.66] &[-0.54, 0.54] &[-0.47, 0.47] &[-0.42, 0.42] &[-0.38, 0.38]\\ \hline 
  \multirow{2}{*}{11.26} & $SEM$  &3.56 &2.52 &2.06 &1.59 &1.13 &0.8 &0.65 &0.5 &0.36 & 0.29 & 0.25 &0.23 &0.21\\  
  & $CI$ &[-6.98, 6.98] &[-4.93, 4.93] &[-4.03, 4.03] &[-3.12, 3.12] &[-2.21, 2.21] &[-1.56, 1.56] &[-1.27, 1.27] &[-0.99, 0.99] &[-0.7, 0.7] & [-0.57, 0.57] &[-0.49, 0.49] &[-0.44, 0.44] &[-0.4, 0.4]\\ \hline 
  \multirow{2}{*}{12} & $SEM$  &3.79 &2.68 &2.19 &1.7 &1.2 &0.85 &0.69 &0.54 &0.38 &0.31 &0.27 &0.24 &0.22\\  
  & $CI$ &[-7.44, 7.44] &[-5.26, 5.26] &[-4.29, 4.29] &[-3.33, 3.33] &[-2.35, 2.35] &[-1.66, 1.66] &[-1.36, 1.36] &[-1.05, 1.05] &[-0.74, 0.74] &[-0.61, 0.61] &[-0.53, 0.53] &[-0.47, 0.47] &[-0.43, 0.43]\\ \hline 
  \multirow{2}{*}{13.12} & $SEM$  &4.15 &2.94 &2.4 &1.86 &1.31 &0.93 &0.76 &0.59 &0.41 &0.34 &0.29 &0.26 &0.24\\  
  & $CI$ &[-8.13, 8.13] &[-5.75, 5.75] &[-4.69, 4.69] &[-3.64, 3.64] &[-2.57, 2.57] &[-1.82, 1.82] &[-1.48, 1.48] &[-1.15, 1.15] &[-0.81, 0.81] &[-0.66, 0.66] &[-0.58, 0.58] &[-0.51, 0.51] &[-0.47, 0.47]\\ \hline 
  \multirow{2}{*}{20} & $SEM$  &6.32 &4.47 &3.65 &2.83 &2.0 &1.41 &1.15 &0.89 &0.63 &0.52 &0.45 &0.4 &0.37\\  
  & $CI$ &[-12.4, 12.4] &[-8.77, 8.77] &[-7.16, 7.16] &[-5.54, 5.54] &[-3.92, 3.92] &[-2.77, 2.77] &[-2.26, 2.26] &[-1.75, 1.75] &[-1.24, 1.24] &[-1.01, 1.01] &[-0.88, 0.88] &[-0.78, 0.78] &[-0.72, 0.72]\\ \hline 
  \multirow{2}{*}{30} & $SEM$  &9.49 &6.71 &5.48 &4.24 &3.0 &2.12 &1.73 &1.34 &0.95 &0.77 &0.67 &0.6 &0.55\\  
  & $CI$ &[-18.59, 18.59] &[-13.15, 13.15] &[-10.74, 10.74] &[-8.32, 8.32] &[-5.88, 5.88] &[-4.16, 4.16] &[-3.39, 3.39] &[-2.63, 2.63] &[-1.86, 1.86] &[-1.52, 1.52] &[-1.31, 1.31] &[-1.18, 1.18] &[-1.07, 1.07]\\ \hline 
  \multirow{2}{*}{50} & $SEM$  &15.81 &11.18 &9.13 &7.07 &5.0 &3.54 &2.89 &2.24 &1.58 &1.29 &1.12 &1.0 &0.91\\  
  & $CI$ &[-30.99, 30.99] &[-21.91, 21.91] &[-17.89, 17.89] &[-13.86, 13.86] &[-9.8, 9.8] &[-6.93, 6.93] &[-5.66, 5.66] &[-4.38, 4.38] &[-3.1, 3.1] &[-2.53, 2.53] &[-2.19, 2.19] &[-1.96, 1.96] &[-1.79, 1.79] \\\hline

  \end{tabular}}
  \end{turn}

  \caption{Table of $\text{SEM}$ and confidence intervals $CI$  for different sizes $k$ of the test set and for different values of $\sigma$ including the experimental values obtained in \tableautorefname~\ref{tab:Stats_on_full_ds}. 
  }
  \label{gaussian}
\end{table*}

\subsection{Dice Similarity Coefficient Performance Measure}

\begin{figure}
    \includegraphics[width=.49\linewidth]{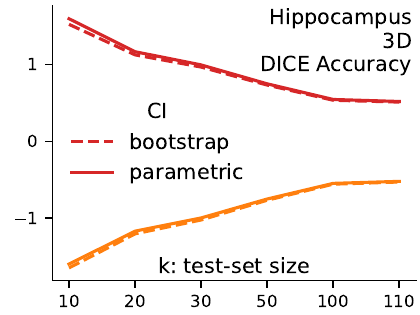}%
    \hfill%
    \includegraphics[width=.49\linewidth]{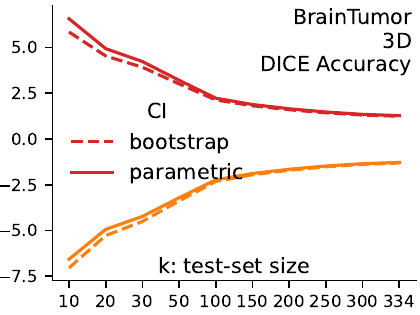}%

    \includegraphics[width=.49\linewidth]{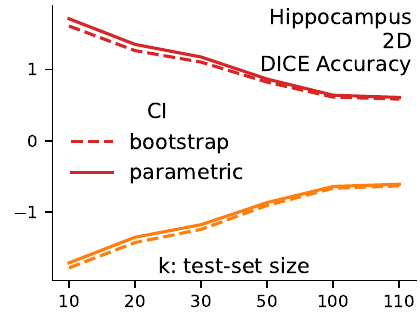}%
    \hfill%
    \includegraphics[width=.49\linewidth]{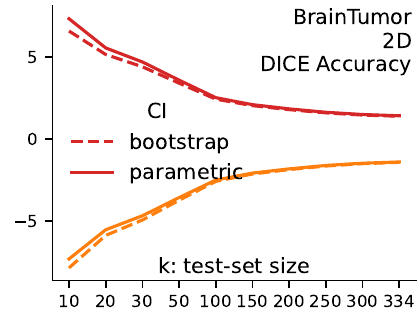}%

    \caption{Parametric and boostrap CIs: Dice Similarity Coefficient (detailed results
    in Tables \ref{table:Hippo-DSC-3D},
	\ref{table:Hippo-DSC-2D},
	\ref{table:Brain-DSC-3D},
	\ref{table:Brain-DSC-2D})}%
    \label{fig:DSC-CIs}%
\end{figure}

Results for the Dice Similarity Coefficient for different sample sizes $k$ are reported
in \autoref{fig:DSC-CIs} (details in Tables~\ref{table:Hippo-DSC-3D},
~\ref{table:Hippo-DSC-2D}, ~\ref{table:Brain-DSC-3D} and
~\ref{table:Brain-DSC-2D}). As expected, as the sample size decreases,
the estimates become less precise (the standard error increases, the
confidence interval widens). Moreover, the confidence intervals obtained
via parametric estimation are very close, though slightly below, to the ones obtained via bootstrapping. Overall, since the CI obtained via the parametric estimates are slightly narrower than that of the bootstap values, one can assume that the parametric estimates could serve as a lower bound to the test set size needed to insure a particular precision. 
Finally, one can observe that bootstrap confidence intervals are slightly skewed. However, this skewness rapidly fades away as the test set sample size $k$ decreases.

\begin{figure}
    \includegraphics[width=.49\linewidth]{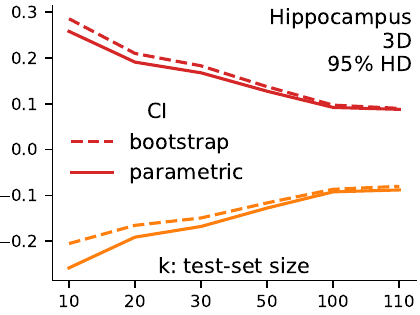}%
    \hfill%
    \includegraphics[width=.49\linewidth]{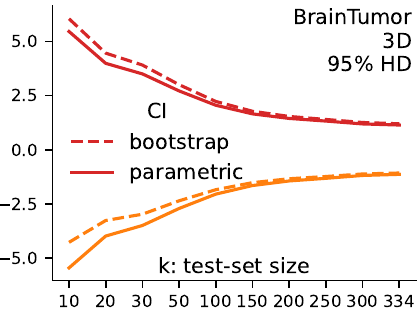}%

    \includegraphics[width=.49\linewidth]{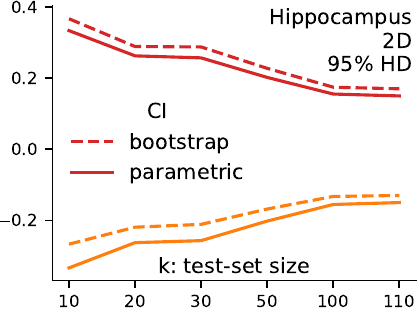}%
    \hfill%
    \includegraphics[width=.49\linewidth]{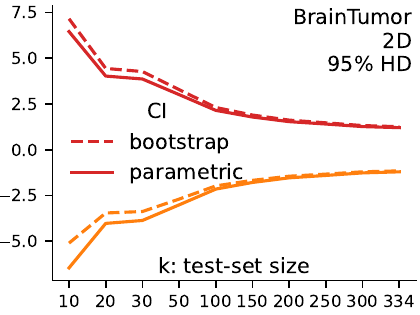}%

    \caption{Parametric and boostrap CIs: Hausdorff Distance (detailed
	results in Tables
	\ref{table:Hippo-HD-3D},
	\ref{table:Hippo-HD-2D},
	\ref{table:Brain-HD-3D},
	\ref{table:Brain-HD-2D}%
    )}%
    \label{fig:HD-CIs}%
\end{figure}

\subsection{Hausdorff Distance Performance Measure}
Results for the Hausdorff distance metric are reported in \autoref{fig:DSC-CIs} (details in Tables~\ref{table:Hippo-HD-3D}, \ref{table:Hippo-HD-2D}, \ref{table:Brain-HD-2D} and \ref{table:Brain-HD-3D}). Even though we saw that the Hausdorff distance is far from being Gaussian distributed, it is interesting to note that obtained parametric estimates are actually close to bootstrap ones. It thus seems reasonable to assume that parametric estimates can, as in the case of the Dice Similarity Coefficient, be used to derive a lower bound on the minimal test set size needed to achieve a given precision. 

Inspecting the normalized width $\nu$ (reported in appendix
\ref{sec:details_results_subsample}) shows that it is much larger for the
Hausdorff distance than for the  Dice Similarity Coefficient for the same model and the
same dataset. This indicates that more test samples are required to achieve a particular precision. Possible explanations include the discontinuity of the Hausdorff distance and the fact that it corresponds to a maximum rather than an average.

\section{A table of Gaussian confidence intervals}
\label{sec:simulation}

The previous section shows that parametric estimates match well
bootstrap confidence intervals. Confidence intervals can be useful even
before running an experiment, to give
a lower bound for the dataset size needed to
achieve a given precision. To facilitate the use of confidence intervals,
Table~\ref{gaussian} list them for different
values of test set size $k$ and $\sigma$ (which reflects how variable is
the performance across test samples), computed via Equations~\ref{eqn:approximation}.
Some values of $\sigma$ match those experimentally measured in our
experiences in Table~\ref{tab:Stats_on_full_ds} (0.47, 0.81, 2.79, 3.26,
10.63, 11.26, 13.12). We recall that the value of $\mu$, in itself, has
no impact on the width of the confidence interval nor on the SEM --even though it is usually observed that lower performing models, thus associated with a lower value of $\mu$, also have a more variable performance and thus a larger value of $\sigma$.

\section{Discussion}
\label{discussion}
We have studied the precision with which segmentation performance can be
estimated in typical 3D brain imaging studies. We provided typical values
for confidence intervals of the performance of trained models. Such
values can be useful to authors and reviewers, \emph{e.g.} to roughly estimate confidence intervals that can be expected for a given study. Finally, we have shown that, under typical performance spreads, the sample size needed to achieve a given confidence interval is smaller than for classification tasks. To that purpose, we have conducted an extensive set of experiments using
both parametric and bootstrap estimates. Results show that parametric
estimates give reasonable results even when the performance metric
substantially deviates from a Gaussian distribution. 

\paragraph*{Test-set size for statistical control}

Classification tasks need large test-set sizes for tight confidence
intervals (though higher-performing models lead to less uncertainty).
Typically, for classification a 1\%-wide confidence interval requires
about $10\,000$ samples for models with over 90\%
accuracy~\cite{varoquaux2022}. 

It is intuitive that the typical test size needed for a given confidence
interval is smaller for segmentation than for classification. This is due to the continuous nature of segmentation metrics which aggregate information across multiple pixels or voxels. On the contrary, in classification, the metric associated to each image is binary (correct or wrong) and these are then aggregated at the sample level to form a continuous metric (e.g. accuracy). In other words, in classification, one image provides only one bit of information to the estimation of the performance. In contrast, in segmentation, each pixel/voxel provides one bit of information, and thus each image contributes many bits. This intuition is indeed verified by our experimental results which show that indeed, to achieve a given CI width, fewer test samples are usually needed for segmentation than for classification. This is
good news, as it is arguably more difficult to obtain a large test set in
the case of segmentation, as it requires voxel-wise annotation by a
trained rater. 

However, {\sl smaller} does not mean {\sl small}.
When dispersion is low ($\sigma \leq 3$ for Dice Similarity Coefficient),
hundreds of samples may suffice for tight confidence intervals (e.g. 1\%-wide). 
When the  dispersion is larger (for example $\sigma \approx
15$), as often for more difficult tasks, a $1\%$-wide
confidence interval may require more than 1\,000 test samples. More relaxed confidence intervals (\emph{e.g.} $4\%$-wide) may still require about 200 test samples.

\begin{figure}[t!]
    \center
    \includegraphics[width=8cm,height=8cm]{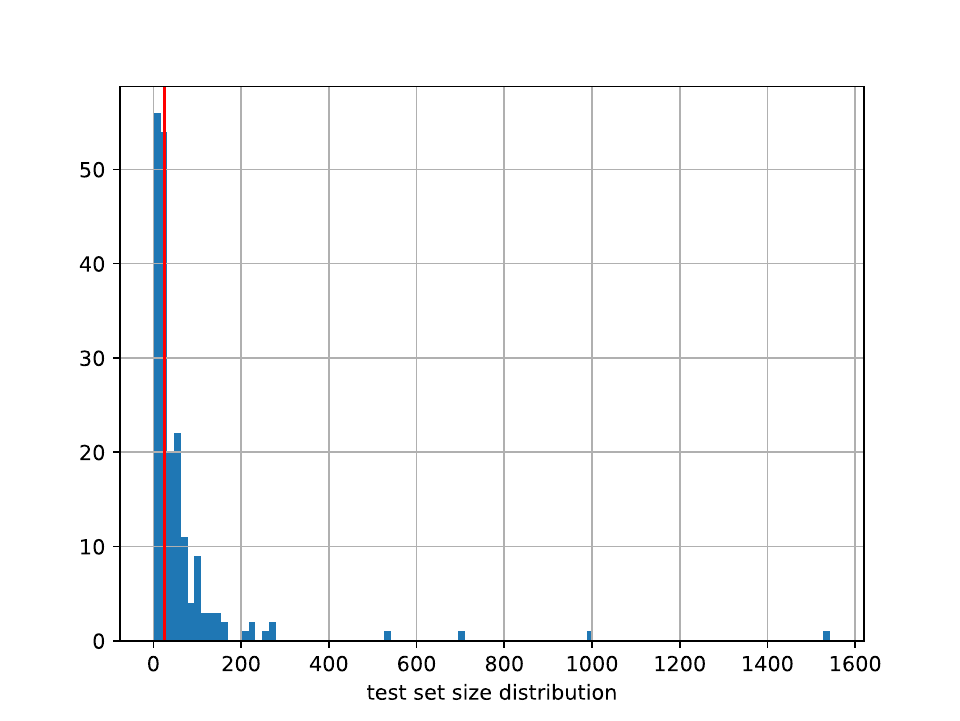} \label{survey}
    \caption{Histogram of the test-set size of different experiments from segmentation papers published in JMI, TMI, and MedIA in 2022. The red line represents the median test size ($n=25$).}
    \label{fig:survey}
\end{figure}


\paragraph*{A call for larger test sets}

Are the practices of medical-imaging studies in line with
such test set sizes? As mentioned in the introduction, we surveyed papers published in 2022 in three medical imaging journals: IEEE Transactions on Medical Imaging (IEEE TMI), Medical Image Analysis (MedIA), and SPIE's Journal of Medical Imaging (JMI) dealing with segmentation of 3D images.
This survey, though not exhaustive, sheds light on current practices in
recent medical-imaging papers.  \figureautorefname~\ref{fig:survey} shows
that the test set size is in general small and highly variable across
studies (median, 25; minimum: 1; maximum: 1543). Thus, with the exception of a few studies, there is a need to increase the size of test sets to obtain reasonably precise estimates of the performance. 

\paragraph*{Reporting confidence intervals is important}
Trustworthy evidence on segmentation tools is important for their
clinical adoption. Confidence intervals and bootstrapping are classic
statistical practice; and yet medical image segmentation papers seldom
report confidence intervals: we found them in only 10\% of papers of our
survey (Table~\ref{tab:survey}).
Beyond immediate clinical adoption, controlled statistical evidence is a component of 
reproducibility, needed for scientific progress~\cite{mcdermott2019reproducibility,colliot2022reproducibility}. 
{\sl Statistical reproducibility} is suited to reason on
generalization to unseen data: results need to 
be statistically compatible rather than identical. Reporting confidence
intervals is central to asserting statistical reproducibility of studies. 
Awareness about this issue has risen, as can been seen for example in
reproducibility checklists (e.g. MICCAI
2021\footnote{\url{https://miccai2021.org/files/downloads/MICCAI2021-Reproducibility-Checklist.pdf}}).
However, the fact that confidence intervals are so rarely reported in
recent papers indicates yet more awareness is needed.

Reporting confidence intervals enables progress at the level of
the community. To disseminate a trained model (for use by clinical
researchers, neuroscientists ...), the users should understand expected
variances in the performance of the model. Confidence intervals also
enable comparing performance to inter-rater (or intra-rater) variability. When computing inter-rater variability on the test set used to evaluate the trained model, one can obtain confidence intervals for the trained model and for the inter-rater and check if they overlap. 

The absence of confidence intervals in model evaluations can have serious implications for the interpretation of published results. When performance metrics such as Dice coefficients or Hausdorff distances are presented without associated confidence intervals, it becomes challenging for the community to assess the reliability of these metrics, especially when comparing different models or methods.  For example, small differences in performance could be due to random fluctuations in the evaluation process rather than genuine improvements. This issue is particularly critical in medical imaging, where decisions based on model outputs can directly impact patient care. Thus, failing to report confidence intervals can lead to overconfidence in model performance and may prevent the detection of models that, while performing well on a given test set, fail to generalize to other datasets or clinical settings.

Confidence intervals (and standard errors) are tools from inferential statistics, specifically from statistical estimation. They indicate how {\sl precise} an {\sl estimate} (e.g. the mean Dice) is. They are dependent on sample size: when the sample size goes to infinity the standard errors tends towards zero and the CI tends towards a single point. Standard-deviation and other quantities such as inter-quartile ranges) are tools from descriptive statistics. They describe how {\sl variable} a {\sl measurement} (e.g. the Dice Similarity Coefficient across the test set) is. They don't shrink with increasing sample size. Of course, the estimate of the standard-deviation is more {\sl precise} for a larger sample but its order of magnitude does not change. Therefore, inferential and descriptive statistics are complementary and both should be reported. 

\paragraph*{A table for confidence intervals}
The width of the confidence interval  depends on two factors: the size of
the test set ($n$) and the variability of the performance ($\sigma$). 
As parametric estimates give reasonable results, one can compute typical confidence intervals for given $n$ and $\sigma$ and we provide this information in Table~\ref{gaussian}.
These values can be useful to researchers preparing a segmentation study,
to help them choose the size of the test set. Investigators must then make an
hypothesis on the expected value of $\sigma$ (as when doing a power
analysis), for example using previously published studies on similar
segmentation tasks (given that these report $\sigma$). The values can
also be useful to reviewers who can use these to gauge
plausible confidence intervals, should these not be
reported in the paper. 

\paragraph*{When standard deviation is not reported}
The width of the confidence interval  depends on two factors: the size of
the test set ($n$) and the variability of the performance ($\sigma$). 
In classification problems, for some metrics such as accuracy, there is a closed-form relationship between SD and the mean since the metric follows a binomial proportion distribution (the variance is $p\times(1-p)$ where $p$ is the accuracy). In such a case,  we have an analytical formula which shows that the variability decreases as performance approaches its extremes. 
As a consequence, for classification, using a binomial distribution gives parametric confidence
intervals~\cite{VAROQUAUX201868,varoquaux2022}. Confidence
intervals can thus easily be computed post-hoc as a function of the accuracy,
\emph{e.g.} when reading a manuscript.

However, for segmentation metrics like the Dice Similarity Coefficient (DSC), no such closed-form relationship exists. Nevertheless, in practice,  \cite{Chr_Confidence_MICCAI2024} have shown that, empirically, one observes the same pattern of relationship between mean and SD, although not in an exact manner. Specifically, they showcase that SD can be approximated by a second-order polynomial function of the mean DSC. Such an approximation can offer a way to  have a rough estimate of SD when it is not reported. This in turns provides a rough confidence interval. This approximation, while useful, does not eliminate the necessity of directly reporting standard-deviation and confidence intervals in future work.

\paragraph*{Preliminary recommendations}
Providing definite recommendations for reporting is clearly beyond the scope of the present paper. Indeed, such recommendations would require further experiments with more tasks, models and metrics as well as involving more members of the community to reach a consensus.

However, we believe that we can still provide preliminary recommendations which would anyway constitute an improvement over current typical practices. We recommend that authors systematically report confidence intervals for trained models. Even though we showed that parametric approximations can yield reasonable estimates, we tend to recommend to use the bootstrap method as it does not make assumptions on the distribution of the metric. Moreover, it is easy to perform since it only requires to resample performance metrics obtained on the test set (without recomputing them). Furthermore, unlike what is commonly believed, bootstrapping on the test set is actually very fast: less than a second on a standard laptop in our experiments (specifically 0.53 second for the hippocampus and 0.62 second for the Brain Tumor dataset). This is negligible compared to the cost of training a deep learning model. Note nevertheless, that both the parametric method and the standard bootstrap assume that the test observations are independent. If this does not hold, CIs will be unvalid. The case of dependent observations is beyond the scope of the present paper.
Finally, confidence intervals should be complemented by descriptive statistics, both in graphical (e.g.  displaying the distribution of the performance metrics with a box plot) and numerical form (reporting standard deviation, median and inter-quartile range for instance).

\paragraph*{More work to benchmark learning procedures}
We investigated
the sampling noise in the test set, the only source of variance when
evaluating a {\sl trained model}. However
evaluating a {\sl learning procedure} --as when comparing two methods--
comes with other sources of variance such as hyperparameters or random
seeds~\cite{bouthillier2021accounting,varoquaux2022}. Methods to build
confidence intervals for {\sl learning procedures} typically involve
running multiple trainings  with different data splits (to account for
sampling noise in training and testing sets) but also different
hyperparameters or random seeds
\cite{bouthillier2021accounting,varoquaux2022}. They require much more
computing power and lead to wider confidence intervals.

\paragraph*{Limitations} While we studied the two most widely used
performance metrics, there are many other metrics, some being more appropriate for
certain settings~\cite{metrics-reloaded-2}. Also, our scope is limited to brain MRI tasks, specifically hippocampus and tumor segmentation. We have included both a relatively easy and a more difficult segmentation task which allows to see the behavior of CIs across different levels of performance. However, other other modalities, such as abdominal CT or cardiac ultrasound, introduce additional variability and annotation challenges. It remains to be seen if our observations hold for other organs and modalities. Also, we only studied one method, the nnU-net, and future work would need to study other architectures.

Furthermore, many factors contribute to the variability of performance, including lesion size and image quality. Larger lesions generally result in higher DSC scores, while smaller lesions often lead to lower DSC scores. Similarly, image quality, which can vary due to factors such as motion artifacts, noise, or differences in resolution, introduces additional variability, making it harder for models to consistently segment lesions accurately across different scans. Assessing the influence of these factors is another important future work.

\paragraph*{Distribution shifts during deployment}
It is crucial to acknowledge the potential for distribution shifts during deployment, such as when applying the method to healthy scans or cases with lesions of significantly different sizes. Confidence intervals provide a measure of variability under random sampling from a given test distribution, but they do not account for distribution shifts that may arise during deployment. Distribution shifts are a separate and important topic, as explored in studies \cite{Koch2024}.

\section{Conclusion}
\label{conclusion}

Confidence intervals are needed to characterize the uncertainty on the
performance of image segmentation tools. 
In the context of 3D brain MRI segmentation, we have shown that simple parametric estimates can serve as a reasonable
reference to guide researchers on the test set size needed for a target
precision. Importantly, results show that the test set size needed to achieve a given precision is  lower for segmentation than for classification tasks. 
Confidence intervals can be computed easily; we hope that they will be
used more in image segmentation studies.

\section*{Disclosure of interests}
{\bf Competing financial interests related to the present article:}
 none to disclose for all authors.
 
{\bf Competing financial interests unrelated to the present article:} OC reports having received consulting fees from AskBio and Therapanacea. OC reports that his laboratory has received grants (paid to the institution) from Qynapse and that members from his laboratory have co-supervised a PhD thesis with Qynapse. OC’s spouse was an employee  myBrainTechnologies and is now an employee of DiamPark. OC holds a patent registered at the International Bureau of the World Intellectual Property Organization (PCT/IB2016/0526993, Schiratti J-B, Allassonniere S, Colliot O, Durrleman S, A method for determining the temporal progression of a biological phenomenon and associated methods and devices) (2017).

\section*{Acknowledgements}

The research leading to these results has received funding from the French government under management of Agence Nationale de la Recherche as part of the “France 2030” program (reference ANR-23-IACL-0008, project PRAIRIE-PSAI), as part of the "Investissements d'avenir" program (reference ANR-19-P3IA-0001, project PRAIRIE 3IA Institute and reference ANR-10-IAIHU-06, project Agence Nationale de la Recherche-10-IA Institut Hospitalo-Universitaire-6), from Agence Nationale de la Recherche (reference ANR-23-CE17-0054, project LYMP-PD), from the European Union’s Horizon Europe Framework Programme (grant number 101136607, project CLARA), was supported by a grant from Inserm and the French Ministry of Health in the context of the MESSIDORE 2023 call operated by IReSP (reference AAP-2023-MSDR-341011),  and by the Fonds Recherche Neurosciences under the project “NEIMO - Monitoring Neuroinflammation: a Paris Brain Institute-Yale Neurology Collaborative program.

\appendix

\subsection{Details on the estimated measures on subsamples of the test sets}%
\label{sec:details_subsample}

\paragraph{Parametric estimates}
For the parametric estimation, we denote as 
\begin{equation*}
    \mu_{k,j}=\frac{1}{k} \sum_{l=1}^{k}P_{j,k,l}, \quad \sigma_{k,j}=\sqrt{\frac{1}{k} \sum_{l=1}^{k}{\left(P_{j,k,l}-\mu_{k,j}\right)^2}}
\end{equation*}
the empirical mean and standard deviation for the subsample $S_{k,j}$ where $P_{j,k,l}$ is the performance metric (in our case either Dice Similarity Coefficient or Hausdorff distance) of a given subject in the sub-sample $S_{k,j}$. Similarly, we use the notations 
\begin{equation*}
    \text{SEM}_{k,j}=\frac{\sigma_{k,j}}{\sqrt{k}}, \quad  \nu_{k,j}=\frac{2*1.96*\text{SEM}_{k,j}}{\mu_{k,j}}.
\end{equation*}
We then compute the average of $\mu_{k,j}$, $\sigma_{k,j}$, $\text{SEM}_{k,j}$, $\nu_{k,j}$ across the different subsamples $S_{k,j}$ for $k$ fixed and  $j \in \{1,\ldots, 100\}$. This provides the following  estimates $\mu_{k}$, $\sigma_{k}$, $\text{SEM}_{k}$ and $\nu_{k}$. Similarly, we define $\text{CI}_{k}=[-1.96\times\text{SEM}_{k},1.96\times\text{SEM}_{k}]$.

\paragraph{Bootstrap estimates}
Bootstrap estimations are performed as follows. For a given subsample
$S_{k,j}$ of size $k$ and index $j$, $M=15\,000$ bootstrap samples of size $k$ are drawn with replacement. We denote a given bootstrap sample as $S^*_{k,j,m}$ and its mean as $\mu^*_{k,j,m}$ where $m \in \{ 1, \ldots, M\}$ is the index of the $m^{th}$ bootstrap sample of subsample $S_{k,j}$. The bootstrap mean $\mu^*_{k,j}$ of $S_{k,j}$ is the mean of the bootstrap sample means   $\mu^*_{k,j,m}$: $$ \mu^*_{k,j} = \frac{1}{M} \sum_{m=1}\mu^*_{k,j,m} $$ The standard error of the mean $\mu^*_{k,j}$ (denoted as $\text{SEM}^*_{k,j}$) obtained via bootstrapping  is the standard deviation of the means of all bootstrap samples of subsample $S_{k,j}$: $$
\text{SEM}^*_{k,j} = \sqrt{\frac{1}{M} \sum_{m=1}^{M}{\left(\mu^*_{k,j,m}-\mu^*_{k,j}\right)^2}}.
$$ The $95\%$ confidence interval 
of the sample $S_{k,j}$ is denoted as $[a^*_{k,j}, b^*_{k,j}]$ and is the set of values between the $2.5 \% $ and $97.5\%$ percentile of the sorted bootstrap means of subsample $S_{k,j}$. 
We then compute the average of $\mu^*_{k,j}$, and $\text{SEM}^*_{k,j}$  across the different subsamples $S_{k,j}$ for $k$ fixed and  $ j \in \{1, ..., 100\}$ and denote the results as $\mu^*_{k}$ and $\text{SEM}^*_{k}$. For the confidence intervals, we compute the average values $a^*_{k}=\frac{1}{100}\sum_{j=1}^{100}{a^*_{k,j}}$ and $b^*_{k}=\frac{1}{100}\sum_{j=1}^{100}{b^*_{k,j}}$ to obtain the average confidence interval independent of the mean $\text{CI}^*_k=[a^*_{k}-\mu^*_{k},b^*_{k}-\mu^*_{k}]$. We finally compute the normalized width of the confidence interval as: $\nu^*_{k}=\frac{b^*_{k}-a^*_{k}}{\mu^*_{k}}$.

\subsection{Detailed results on subsamples of the test sets}
\label{sec:details_results_subsample}

Detailed results on subsamples of the test set, which were summarized in Figures~\ref{fig:DSC-CIs} and~\ref{fig:HD-CIs}, are presented in Tables~\ref{table:Hippo-DSC-3D}, \ref{table:Hippo-DSC-2D}, \ref{table:Brain-DSC-3D}, \ref{table:Brain-DSC-2D}, \ref{table:Hippo-HD-3D}, \ref{table:Hippo-HD-2D}, \ref{table:Brain-HD-3D} and \ref{table:Brain-HD-2D}.

\begin{table*}[b]
	\begin{center}
		\begin{tabular}{|l|ccccc|cccc|}
            \hline
            &  \multicolumn{5}{c|}{Parametric} & \multicolumn{4}{c|}{Bootstrap} \\
			Subsample size $k$ & $\mu_k$ & $\sigma_k$ & $SEM_k$ & $CI_k$ & $\nu_k$ & $\mu ^*_k$ & $SEM^*_k$ & $CI^*_k $ & $\nu^*_k$ \\
			\hline
			10 & $89.751 $  & $2.578 $ & $0.815$  & [-1.6, 1.6] & 0.036 & $89.752 $  & $0.815$ & $[-1.647, 1.525]$ & 0.035 \\
			20 & $89.723 $  & $2.666 $ & $0.596$  & [-1.17, 1.17] & 0.026 & $89.724 $  & $0.597$ & $[-1.204, 1.128]$ & 0.026 \\
			30 & $89.681 $  & $2.785 $ & $0.508$  & [-0.995, 0.995] & 0.022 & $89.682 $  & $0.508$ & $[-1.023, 0.968]$ & 0.022 \\
			50 & $89.768 $  & $2.707 $ & $0.383$  & [-0.75, 0.75] & 0.017 & $89.768 $  & $0.383$ & $[-0.766, 0.734]$ & 0.017 \\
			100 & $89.721 $  & $2.788 $ & $0.279$  & [-0.545, 0.545] & 0.012 & $89.721 $  & $0.279$ & $[-0.557, 0.536]$ & 0.012 \\
			110 & $89.714$  & $2.784 $ & $0.265$  & [-0.52, 0.52] & 0.012 & $89.714 $  & $0.266$ & $[-0.529, 0.512]$ & 0.012 \\\hline
		\end{tabular}%
	\end{center}
	\caption{Hippocampus dataset, 3D U-net, Dice accuracy. Results on sub-samples of size $k \leq 110$. We show the mean over all the sub-samples $S_{k,j}$ of a given size $k$ ($k$ is fixed and $j \in \{1, \ldots, 100\}$).}
  \vspace{1cm}
	\label{table:Hippo-DSC-3D}
\end{table*}

\begin{table*}[b]
	\begin{center}
		\begin{tabular}{|l|ccccc|cccc|}
  \hline
            &  \multicolumn{5}{c|}{Parametric} & \multicolumn{4}{c|}{Bootstrap} \\			Subsample size $k$ & $\mu_k$ & $\sigma_k$ & $SEM_k$ & $CI_k$ & $\nu_k$ & $\mu ^*_k$ & $SEM^*_k$ & $CI^*_k $ & $\nu^*_k$ \\
			\hline
			10 & $88.334 $  & $2.759 $ & $0.872$  & [-1.71, 1.71] & 0.039 & $88.335 $  & $0.872$ & $[-1.78, 1.608]$ & 0.038 \\
			20 & $88.189 $  & $3.082 $ & $0.689$  & [-1.35, 1.35] & 0.031 & $88.19 $  & $0.689$ & $[-1.425, 1.263]$ & 0.030 \\
			30 & $88.126 $  & $3.281 $ & $0.599$  & [-1.175, 1.175] & 0.027 & $88.126 $  & $0.599$ & $[-1.24, 1.102]$ & 0.027 \\
			50 & $88.254 $  & $3.12 $ & $0.441$  & [-0.865, 0.865] & 0.020 & $88.255 $  & $0.441$ & $[-0.905, 0.822]$ & 0.020 \\
			100 & $88.212 $  & $3.261 $ & $0.326$  & [-0.64, 0.64] & 0.015 & $88.212 $  & $0.326$ & $[-0.664, 0.612]$ & 0.014 \\
			110 & $88.197$  & $3.252 $ & $0.31$  & [-0.61, 0.61] & 0.014 & $88.197 $  & $0.31$ & $[-0.631, 0.584]$ & 0.014 \\ \hline
		\end{tabular}
 
	\end{center}
	\caption{Hippocampus dataset, 2D U-net, Dice Accuracy. Results on subsamples of size $k \leq 110$. We show the mean over all the sub-samples $S_{k,j}$ of a given size $k$ ($k$ is fixed and $j \in \{1, \ldots, 100\}$).}
 \vspace{1cm}
	\label{table:Hippo-DSC-2D}
\end{table*}

\begin{table*}[b]
	\begin{center}
		\begin{tabular}{|l|ccccc|cccc|}
                \hline
            &  \multicolumn{5}{c|}{Parametric} & \multicolumn{4}{c|}{Bootstrap} \\			Subsample size $k$ & $\mu_k$ & $\sigma_k$ & $SEM_k$ & $CI_k$ & $\nu_k$ & $\mu ^*_k$ & $SEM^*_k$ & $CI^*_k $ & $\nu^*_k$ \\
			\hline
			10 & $80.457 $  & $10.612 $ & $3.356$  & [-6.58, 6.58] & 0.164 & $80.461$  & $3.356$ & $[-7.051, 5.853]$ & 0.160 \\
			20 & $80.188 $  & $11.281 $ & $2.522$  & [-4.945, 4.945] & 0.123 & $80.189 $  & $2.523$ & $[-5.28, 4.541]$ & 0.122 \\
			30 & $79.925 $  & $11.839 $ & $2.162$  & [-4.235, 4.235] & 0.106 & $79.924 $  & $2.162$ & $[-4.512, 3.925]$ & 0.106 \\
			50 & $80.375 $  & $11.593 $ & $1.64$  & [-3.215, 3.215] & 0.080 & $80.374 $  & $1.639$ & $[-3.384, 3.015]$ & 0.080 \\
			100 & $80.388 $  & $11.389 $ & $1.139$  & [-2.23, 2.23] & 0.055 & $80.388 $  & $1.139$ & $[-2.321, 2.137]$ & 0.055 \\
			150 & $80.329 $  & $11.788 $ & $0.962$  & [-1.885, 1.885] & 0.047 & $80.328 $  & $0.963$ & $[-1.954, 1.817]$ & 0.047 \\
			200 & $80.253 $  & $11.936 $ & $0.844$  & [-1.655, 1.655] & 0.041 & $80.252 $  & $0.844$ & $[-1.709, 1.601]$ & 0.041 \\
			250 & $80.295 $  & $11.919 $ & $0.754$  & [-1.475, 1.475] & 0.037 & $80.296 $  & $0.754$ & $[-1.521, 1.431]$ & 0.037 \\
			300 & $80.28 $  & $11.924 $ & $0.688$  & [-1.35, 1.35] & 0.034 & $80.281 $  & $0.688$ & $[-1.386, 1.311]$ & 0.034 \\
			334 & $80.265$  & $11.929 $ & $0.653$  & [-1.28, 1.28] & 0.032 & $80.264 $  & $0.653$ & $[-1.313, 1.245]$ & 0.032 \\ \hline
		\end{tabular}
	\end{center}
	\caption{Brain Tumor dataset, 3D U-net, Dice Accuracy. Results on subsamples of size $k \leq 334$.  We show the mean over all the sub-samples $S_{k,j}$ of a given size $k$ ($k$ is fixed and $j \in \{1, \ldots, 100\}$).}
  \vspace{2cm}
	\label{table:Brain-DSC-3D}
\end{table*}

\begin{table*}[b]
	\begin{center}
 
		\begin{tabular}{|l|ccccc|cccc|}
  \hline
            &  \multicolumn{5}{c|}{Parametric} & \multicolumn{4}{c|}{Bootstrap} \\			Subsample size $k$ & $\mu_k$ & $\sigma_k$ & $SEM_k$ & $CI_k$ & $\nu_k$ & $\mu ^*_k$ & $SEM^*_k$ & $CI^*_k $ & $\nu^*_k$ \\
			\hline
			10 & $77.811 $  & $11.799 $ & $3.731$  & [-7.315, 7.315] & 0.188 & $77.815 $  & $3.731$ & $[-7.856, 6.566]$ & 0.185 \\
			20 & $77.334 $  & $12.631 $ & $2.824$  & [-5.535, 5.535] & 0.143 & $77.334 $  & $2.825$ & $[-5.879, 5.137]$ & 0.142 \\
			30 & $77.101 $  & $13.084 $ & $2.389$  & [-4.68, 4.68] & 0.121 & $77.101 $  & $2.389$ & $[-4.94, 4.389]$ & 0.121 \\
			50 & $77.607 $  & $12.897$ & $1.824$  & [-3.575, 3.575] & 0.092 & $77.606 $  & $1.823$ & $[-3.736, 3.39]$ & 0.092 \\
			100 & $77.564 $  & $12.797 $ & $1.28$  & [-2.51, 2.51] & 0.065 & $77.564 $  & $1.279$ & $[-2.588, 2.42]$ & 0.065 \\
			150 & $77.583 $  & $13.01 $ & $1.062$  & [-2.08, 2.08] & 0.054 & $77.582 $  & $1.063$ & $[-2.139, 2.023]$ & 0.054 \\
			200 & $77.468 $  & $13.182 $ & $0.932$  & [-1.825, 1.825] & 0.047 & $77.467 $  & $0.932$ & $[-1.871, 1.778]$ & 0.047 \\
			250 & $77.549 $  & $13.056 $ & $0.826$  & [-1.62, 1.62] & 0.042 & $77.549 $  & $0.826$ & $[-1.654, 1.582]$ & 0.042 \\
			300 & $77.501 $  & $13.104 $ & $0.757$  & [-1.485, 1.485] & 0.038 & $77.502 $  & $0.756$ & $[-1.511, 1.448]$ & 0.038 \\
			334 & $77.489$  & $13.115 $ & $0.718$  & [-1.405, 1.405] & 0.036 & $77.488 $  & $0.717$ & $[-1.43, 1.376]$ & 0.036 \\\hline
		\end{tabular}
	\end{center}
	\caption{Brain Tumor dataset, 2D U-net, Dice Accuracy. Results on subsamples of size $k \leq 334$. We show the mean over all the sub-samples $S_{k,j}$ of a given size $k$ ($k$ is fixed and $j \in \{1, \ldots, 100\}$).}
	\label{table:Brain-DSC-2D}
\end{table*}

\begin{table*}[b]
	\begin{center}
		\begin{tabular}{|l|ccccc|cccc|}
            \hline
            &  \multicolumn{5}{c|}{Parametric} & \multicolumn{4}{c|}{Bootstrap} \\			Subsample size $k$ & $\mu_k$ & $\sigma_k$ & $SEM_k$ & $CI_k$ & $\nu_k$ & $\mu ^*_k$ & $SEM^*_k$ & $CI^*_k $ & $\nu^*_k$ \\
			\hline
			10 & $1.214 $  & $0.416 $ & $0.132$  & [-0.26, 0.26] & 0.428 & $1.214 $  & $0.131$ & $[-0.205, 0.286]$ & 0.404 \\
			20 & $1.206 $  & $0.435 $ & $0.097$  & [-0.19, 0.19] & 0.315 & $1.206 $  & $0.097$ & $[-0.165, 0.21]$ & 0.310 \\
			30 & $1.221 $  & $0.468 $ & $0.086$  & [-0.17, 0.17] & 0.278 & $1.22 $  & $0.086$ & $[-0.149, 0.183]$ & 0.271 \\
			50 & $1.203 $  & $0.46 $ & $0.065$  & [-0.125, 0.125] & 0.208 & $1.203 $  & $0.065$ & $[-0.116, 0.137]$ & 0.211 \\
			100 & $1.205 $  & $0.47 $ & $0.047$  & [-0.09, 0.09] & 0.149 & $1.205 $  & $0.047$ & $[-0.086, 0.097]$ & 0.152 \\
			110 & $1.205$  & $0.47 $ & $0.045$  & [-0.09, 0.09] & 0.149 & $1.205$  & $0.045$ & $[-0.08, 0.09]$ & 0.149 \\\hline
		\end{tabular}
	\end{center}
	\caption{Hippocampus dataset, 3D U-net, Hausdorff Distance. Results on sub-samples of size $k \leq 110$. We show the mean over all the sub-samples $S_{k,j}$ of a given size $k$ ($k$ is fixed and $j \in \{1, \ldots, 100\}$).}
   \vspace{1cm}
	\label{table:Hippo-HD-3D}
\end{table*}

\begin{table*}[b]
	\begin{center}
		\begin{tabular}{|l|ccccc|cccc|}
            \hline
            &  \multicolumn{5}{c|}{Parametric} & \multicolumn{4}{c|}{Bootstrap} \\			Subsample size $k$ & $\mu_k$ & $\sigma_k$ & $SEM_k$ & $CI_k$ & $\nu_k$ & $\mu ^*_k$ & $SEM^*_k$ & $CI^*_k $ & $\nu^*_k$ \\
			\hline
			10 & $1.296$  & $0.538 $ & $0.17$  & [-0.335, 0.335] & 0.517 & $1.296 $  & $0.17$ & $[-0.267, 0.367]$ & 0.490 \\
			20 & $1.302 $  & $0.599 $ & $0.134$  & [-0.265, 0.265] & 0.407 & $1.302 $  & $0.134$ & $[-0.219, 0.289]$ & 0.391 \\
			30 & $1.331 $  & $0.718 $ & $0.131$  & [-0.255, 0.255] & 0.383 & $1.331 $  & $0.131$ & $[-0.211, 0.288]$ & 0.375 \\
			50 & $1.31 $  & $0.729 $ & $0.103$  & [-0.2, 0.2] & 0.305 & $1.31 $  & $0.103$ & $[-0.168, 0.228]$ & 0.303 \\
			100 & $1.31 $  & $0.793 $ & $0.079$  & [-0.155, 0.155] & 0.237 & $1.31 $  & $0.079$ & $[-0.133, 0.175]$ & 0.234 \\
			110 & $1.311$  & $0.803 $ & $0.077$  & [-0.15, 0.15] & 0.229 & $1.311 $  & $0.077$ & $[-0.13, 0.17]$ & 0.229 \\ \hline
		\end{tabular}
	\end{center}
	\caption{Hippocampus dataset, 2D U-net, Hausdorff Distance. Results on subsamples of size $k \leq 110$. We show the mean over all the sub-samples $S_{k,j}$ of a given size $k$ ($k$ is fixed and $j \in \{1, \ldots, 100\}$).}
	\label{table:Hippo-HD-2D}
\end{table*}

\begin{table*}[b]
	\begin{center}
		\begin{tabular}{|l|ccccc|cccc|}
            \hline
            &  \multicolumn{5}{c|}{Parametric} & \multicolumn{4}{c|}{Bootstrap} \\			Subsample size $k$ & $\mu_k$ & $\sigma_k$ & $SEM_k$ & $CI_k$ & $\nu_k$ & $\mu ^*_k$ & $SEM^*_k$ & $CI^*_k $ & $\nu^*_k$ \\
			\hline
			10 & $8.171 $  & $8.806 $ & $2.785$  & [-5.46, 5.46] & 1.336 & $8.168 $  & $2.784$ & $[-4.278, 6.046]$ & 1.264 \\
			20 & $7.635 $  & $9.098 $ & $2.034$  & [-3.985, 3.985] & 1.044 & $7.634 $  & $2.034$ & $[-3.268, 4.459]$ & 1.012 \\
			30 & $7.95 $  & $9.782 $ & $1.786$  & [-3.5, 3.5] & 0.881 & $7.95 $  & $1.787$ & $[-2.98, 3.909]$ & 0.866 \\
			50 & $7.563 $  & $9.791 $ & $1.385$  & [-2.715, 2.715] & 0.718 & $7.563 $  & $1.383$ & $[-2.36, 3.002]$ & 0.709 \\
			100 & $7.702 $  & $10.468 $ & $1.047$  & [-2.05, 2.05] & 0.532 & $7.702 $  & $1.046$ & $[-1.848, 2.228]$ & 0.529 \\
			150 & $7.691 $  & $10.339 $ & $0.844$  & [-1.655, 1.655] & 0.430 & $7.693 $  & $0.844$ & $[-1.522, 1.775]$ & 0.428 \\
			200 & $7.698 $  & $10.413 $ & $0.736$  & [-1.445, 1.445] & 0.375 & $7.697 $  & $0.736$ & $[-1.341, 1.535]$ & 0.374 \\
			250 & $7.758 $  & $10.663 $ & $0.674$  & [-1.32, 1.32] & 0.340 & $7.757 $  & $0.674$ & $[-1.238, 1.397]$ & 0.340 \\
			300 & $7.695 $  & $10.513 $ & $0.607$  & [-1.19, 1.19] & 0.309 & $7.694 $  & $0.607$ & $[-1.12, 1.257]$ & 0.309 \\
			334 & $7.726$  & $10.618 $ & $0.581$  & [-1.14, 1.14] & 0.295 & $7.726 $  & $0.581$ & $[-1.076, 1.196]$ & 0.294 \\\hline
		\end{tabular}
	\end{center}
	\caption{Brain Tumor dataset, 3D U-net, Hausdorff Distance. Results on subsamples of size $k \leq 334$. We show the mean over all the sub-samples $S_{k,j}$ of a given size $k$ ($k$ is fixed and $j \in \{1, \ldots, 100\}$).}
	\label{table:Brain-HD-3D}
\end{table*}

\begin{table*}[b]
	\begin{center}
		\begin{tabular}{|l|ccccc|cccc|}
  \hline
            &  \multicolumn{5}{c|}{Parametric} & \multicolumn{4}{c|}{Bootstrap} \\			Subsample size $k$ & $\mu_k$ & $\sigma_k$ & $SEM_k$ & $CI_k$ & $\nu_k$ & $\mu ^*_k$ & $SEM^*_k$ & $CI^*_k $ & $\nu^*_k$ \\
			\hline
			10 & $9.544 $  & $10.409 $ & $3.292$  & [-6.45, 6.45] & 1.352 & $9.544 \pm 3.441$  & $3.294$ & $[-5.099, 7.152]$ & 1.284 \\
			20 & $8.454 $  & $9.178 $ & $2.052$  & [-4.02, 4.02] & 0.951 & $8.453 \pm 2.302$  & $2.05$ & $[-3.454, 4.441]$ & 0.934 \\
			30 & $9.204 $  & $10.794 $ & $1.971$  & [-3.86, 3.86] & 0.839 & $9.205 \pm 1.786$  & $1.973$ & $[-3.374, 4.269]$ & 0.830 \\
			50 & $8.836 $  & $10.86 $ & $1.536$  & [-3.01, 3.01] & 0.681 & $8.837 \pm 1.359$  & $1.537$ & $[-2.696, 3.282]$ & 0.676 \\
			100 & $8.758$  & $10.966 $ & $1.097$  & [-2.15, 2.15] & 0.491 & $8.756 \pm 0.938$  & $1.096$ & $[-1.974, 2.307]$ & 0.489 \\
			150 & $8.886 $  & $11.158 $ & $0.911$  & [-1.785, 1.785] & 0.402 & $8.885 \pm 0.673$  & $0.912$ & $[-1.671, 1.894]$ & 0.401 \\
			200 & $8.806 $  & $11.068 $ & $0.783$  & [-1.535, 1.535] & 0.349 & $8.806 \pm 0.471$  & $0.783$ & $[-1.445, 1.617]$ & 0.348 \\
			250 & $8.873 $  & $11.32 $ & $0.716$  & [-1.405, 1.405] & 0.317 & $8.873 \pm 0.357$  & $0.715$ & $[-1.33, 1.469]$ & 0.315 \\
			300 & $8.83 $  & $11.173 $ & $0.645$  & [-1.265, 1.265] & 0.287 & $8.831 \pm 0.214$  & $0.645$ & $[-1.204, 1.32]$ & 0.286 \\
			334 & $8.855$  & $11.245 $ & $0.615$  & [-1.205, 1.205] & 0.272 & $8.856 $  & $0.615$ & $[-1.154, 1.257]$ & 0.272 \\ \hline
		\end{tabular}
	\end{center}
	\caption{Brain Tumor dataset, 2D U-net, Hausdorff Distance. Results on sub-samples of size $k \leq 334$. We show the mean over all the sub-samples $S_{k,j}$ of a given size $k$ ($k$ is fixed and $j \in \{1, \ldots, 100\}$).}
	\label{table:Brain-HD-2D}
\end{table*}

\end{document}